   \documentclass[twocolumn,floats,floatfix,superscriptaddress,aps,pra]{revtex4}

\usepackage{amsfonts}
\usepackage{amssymb}
\usepackage{amsmath}
\usepackage{calc}
\usepackage{graphicx}
\usepackage{bm} 
\usepackage{xcolor} 

\def\be{ \begin{equation} }
\def\ee{ \end{equation} }
\def\bea{ \begin{eqnarray} }
\def\eea{ \end{eqnarray} }
\def\bse{ \begin{subequations} }
\def\ese{ \end{subequations} }

\def\H{\mathbf{H}} 
\def\U{\mathbf{U}} 

\def\c{\mathbf{c}}

\def\e{e}

\def\ii{{\rm i}} 
\def\e{{\rm e}} 
\def\t{(t)} 
\def\H{{\sf W}} 
\def\Ha{{\sf W}^a} 
\def\U{{\sf U}} 
\def\bM{\left[\begin{array}{ccccccc}} 
\def\eM{\end{array}\right]} 
\def\ddt{\frac{d}{dt}} 

\def\H{{\sf H}} 
\def\Ha{{\sf H}_a} 
\def\H{{\mathbf H}} 
\def\Ha{{\mathbf H}_a} 
\def\U{{\mathbf U}} 

\def\eigen{\epsilon} 
\def\calA{\mathcal{A}} 

\def\to{\rightarrow}

\def\half{\tfrac12}

\def\ket#1{\vert #1 \rangle}
\def\bra#1{\langle #1 \vert}

\def\endit{ \end{document} }

 \def\rd{\color{red}} 

\def\bws#1{{\rd #1}}

\begin{document}

\author{Boyan T. Torosov}
\affiliation{Institute of Solid State Physics, Bulgarian Academy of Sciences, 72 Tsarigradsko chauss\'{e}e, 1784 Sofia, Bulgaria}
\author{Bruce W. Shore}
\affiliation{618 Escondido Circle, Livermore, California 94550, USA}
\author{Nikolay V. Vitanov}
\affiliation{Department of Physics, Sofia University, 5 James Bourchier blvd, 1164 Sofia, Bulgaria}

\title{Coherent control techniques for two-state quantum systems: A comparative study}

\date{\today}

\begin{abstract}
We evaluate various sources of errors that occur when attempting to produce a specified coherent change of a two-state quantum system using six popular coherent control techniques: resonant excitation, adiabatic following, composite adiabatic passage, universal composite pulses, shortcut to adiabaticity, and single-shot shaped pulses.
As error sources we consider spatial intensity distribution, transit time variation, inhomogeneous broadening, Doppler broadening, unwanted chirp and shape errors.
For the various error types different techniques emerge as the best performers but overall, we find that universal composite pulses
perform most consistently and are most resilient to errors compared to all other procedures.
\end{abstract}

\maketitle



\section{Introduction}

The two-state model of a simple coherently-driven quantum system has long been a fixture of many branches of physics that treat coherent changes of quantum systems \cite{Allen1975}.
Notably,  it has been used in treatments of coherent atomic excitation \cite{Shore1990}, nuclear magnetic resonance (NMR) \cite{NMR}, quantum information \cite{qinfo},  quantum optics \cite{qoptics}, doped solids \cite{GenovUniversal}, chemical reactions \cite{chemical}, and numerous other areas of experimental physics and chemistry.
Typically the simplicity of the mathematics makes possible either closed-form expressions, or simple simulations,  for quantum-state probabilities  that are taken as guides to more realistic descriptions.
Most often the intent is to devise an experimental procedure, in the form of coherent radiation pulses and controlled environment,  that produces a particularly desirable quantum-state change.
Several classes of pulses have been employed for that purpose, each with its advocates.
Here we compare the advantages and disadvantages of six of the most popular choices  and provide quantitative examples of the limitations one may anticipate with each.


The dynamics of the two probability amplitudes $c_1\t$ and $c_2\t$  that define the two-state system and its probabilities $P_n\t = |c_n\t|^2$ for finding the system to be in state $n$ at time $t$ are governed by the time-dependent Schr\"{o}dinger equation, 
\be\label{Schrodinger}
\ii  \hbar \ddt \c(t) = \H(t)\c(t),
\ee
where the column vector $\c(t) = [c_1(t),c_2(t)]^T$ is the state vector in a rotating reference frame \cite{Shore2011}.
The changes to the state vector described here are driven by a $2 \times 2$ Hamiltonian matrix $\H\t$ which is obtained from a semiclassical Hamiltonian expressed in rotating coordinates and with the rotating-wave approximation (RWA) to give slowly varying elements \cite{Shore2011}.
The matrix $\H\t$, written as
\be\label{Hamiltonian}
\H(t) = \half \hbar
\bM
	-\Delta(t) & \Omega(t) \\  \Omega(t) & \Delta(t)
\eM,
\ee
has two  controls, each of which may be crafted experimentally with quite general time variation:
the Rabi frequency  $\Omega(t)$, describing the strength of the interaction between the quantum system and the external field, and the detuning $\Delta(t)$, which is  the  difference between the Bohr transition frequency $\omega_{12}$ and the carrier frequency $\omega$, each of which may have time variation. In order for the RWA to be valid, the Rabi frequency needs to be much smaller than the transition frequency, $\Omega \ll \omega_{12}$, which is usually fulfilled in most physical systems.



The elements of the two-state propagator corresponding to the traceless Hamiltonian of Eq.~\eqref{Hamiltonian}  are expressed via the two complex-valued Cayley-Klein (CK) parameters $a$ and $b$,
\be\label{Propagator}
\U=\left[\begin{array}{cc}
	a & b \\ - b^\ast & a^\ast
\end{array}\right],
\ee
with $|a|^2 + |b|^2 = 1$.
In particular,  the transition probability from initial state 1 to final state 2  (and {\it vice versa}) is expressible as
$P=  p_{21} = p_{12} = |U_{21}|^2 = |b|^2$.
Analytic expressions for the CK parameters can be obtained for various pulses, as defined by expressions for the time dependence of the  Rabi frequency and the detuning.
We shall be interested in how any uncontrollable and unwanted deviations from the intended forms of these controls alter the propagator --- deviations that we regard as errors.




There are several common protocols for crafting   $\Omega\t$ and $\Delta\t$,
 by means of pulsed coherent radiation such as a laser provides,  to produce specified coherent changes, either of populations (quantum-state probabilities) or of coherent superpositions (coherences).
In this paper we will be examining several of these pulse designs and compare their sensitivity to experimental errors.
The next section discusses examples of several protocols for choice of pulses.

\section{Coherent control techniques}

\subsection{Resonant excitation (RE)}

The most commonly used manipulation technique relies on resonant excitation (RE), in which the carrier frequency of the radiation exactly matches the (constant) Bohr frequency of the transition, meaning $\Delta = 0$.
The transition probability for resonant excitation is well known,
\be	\label{eq-resonantp}
P = \sin^2(\calA/2) = \half [ 1 - \cos(\calA)],
\ee
where
\be
\calA = \int dt \ \Omega\t
\label{eq-area}
\ee
is the (temporal) {\em pulse area}, or also known as the Rabi angle.
By choosing the pulse area to  be any odd-integer multiple of $\pi$ (a $\pi$ pulse) we produce complete \emph{population inversion}, or \emph{population swapping}.
Similarly, a pulse area of $\pi/2$ will produce a superposition of equal magnitudes of the two states.
These results depend only on the pulse area, whatever the details of the temporal pulse shape might be.

The practical limitation  of the resonant method is constrained by the requirement  that  the detuning must vanish and the pulse area must be fixed exactly.
The possible difficulty, or impracticality, of these requirements necessitates alternative designs for the pulse controls.

To be specific, here we consider a Gaussian pulse,
\be\label{Gaussian}
\Omega(t) = \Omega_0 \e^{-(t/T)^2},
\ee
where $\Omega_0$ is the peak Rabi frequency and $T$ is the pulse duration.
On resonance, the transition probability is given by Eq.~\eqref{eq-resonantp} with the pulse area  $\calA=\Omega_0 T\sqrt{\pi}$.
Therefore, for the Gaussian pulse, we achieve complete population transfer for $\Omega_0 = \sqrt{\pi}/T$.
For visualization purpose, we  plot the Rabi frequency and the detuning for all of the studied techniques in Fig.~\ref{fig:shapes}.

\subsection{Adiabatic following (AF) }

\begin{figure}[tb]
	\includegraphics[width=1.\columnwidth]{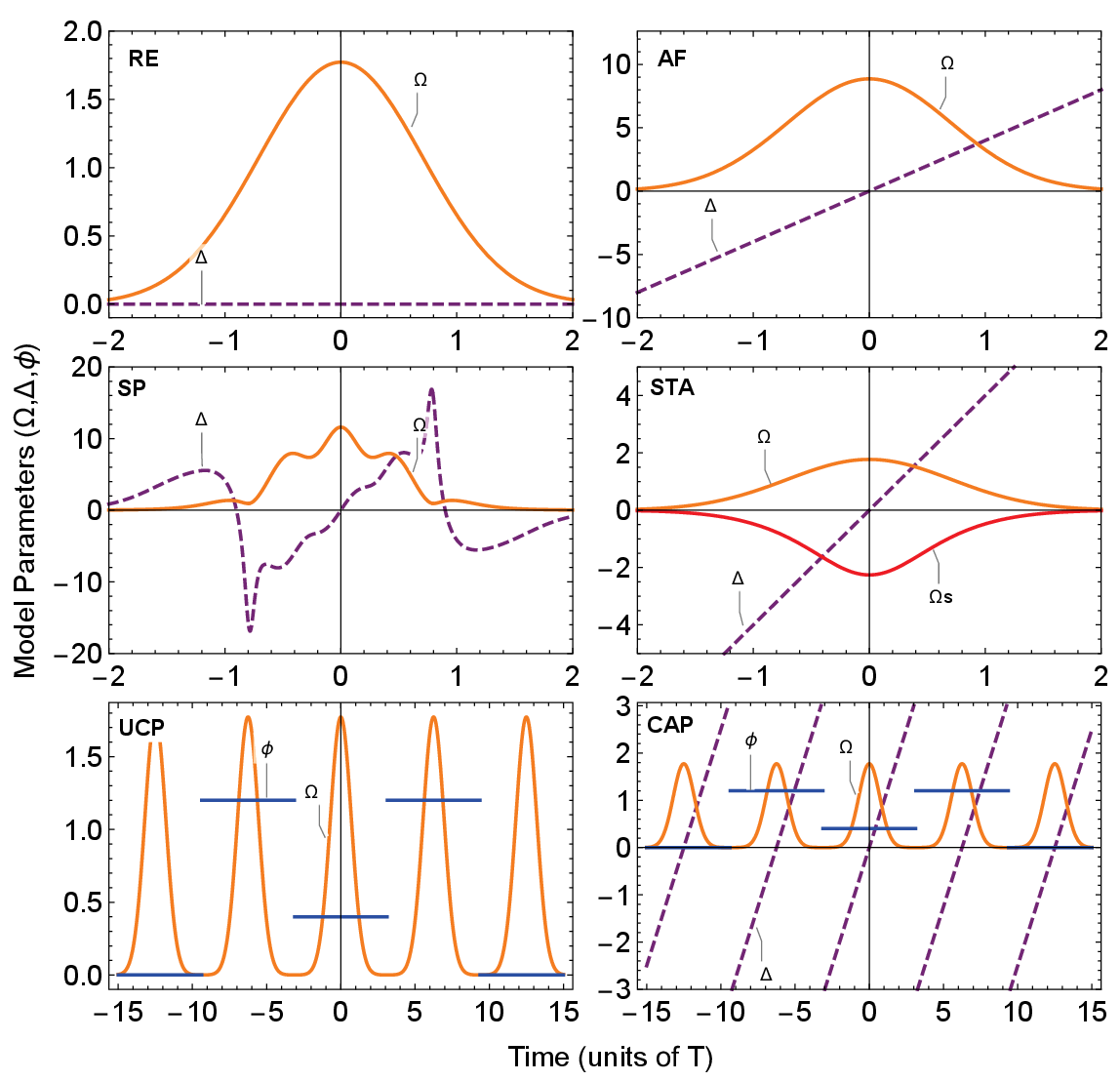}
	\caption{Time dependence of the Rabi frequency and detuning for the six techniques studied here: RE (resonant excitation), AF (adiabatic following), SP (shaped pulses), STA (shortcut to adiabaticity), UCP (universal composite pulses), and CAP (composite adiabatic passage).
}
	\label{fig:shapes}
\end{figure}

A popular procedure for producing population alteration without requiring precise control of the pulse area and the detuning is the {\em adiabatic following} (AF) \cite{Allen1975}.
It relies on matching the initial state vector to one of the two instantaneous eigenvectors of the Hamiltonian.
By adjusting the Hamiltonian time variation to be appropriately slow ({\em adiabatic}) an experimenter can guide the state vector in its Hilbert space to produce any desired superposition state.
In particular, the evolution may produce population swapping, as with a resonant $\pi$ pulse, here termed {\em adiabatic passage}.
To this end, in addition to the Gaussian pulse shape \eqref{Gaussian}, we assume a linearly changing detuning,
\be\label{chirp}
\Delta(t) = \beta t/T ,
\ee
see Fig~\ref{fig:shapes}.
The efficiency of this transfer is limited by the adiabatic condition, which requires slow evolution.
Indeed, the Hamiltonian in the adiabatic basis reads \cite{Vitanov2001}
\be\label{HamiltonianAdiabatic}
\Ha(t) = \hbar \bM
	\epsilon_{-}(t) & -\ii \dot{\theta}(t) \\  \ii \dot{\theta}(t) & \epsilon_{+}(t)
\eM,
\ee
where $ \hbar \eigen_{\pm}(t) = \pm \half \hbar \sqrt{\Omega(t)^2+\Delta(t)^2}$ are the eigenvalues of the Hamiltonian.
 The nonadiabatic coupling (the off-diagonal element) is proportional to the time derivative $\dot{\theta}(t)$ of the mixing angle,
\be
\theta(t)=\half\arctan\left[ \frac{\Omega(t)}{\Delta(t)} \right].
\ee
The state vector will remain aligned with an initial adiabatic state  as long as the change of the angle $\theta(t)$ is much slower than the eigenvalue separation,
 \be
|\dot{\theta}(t)| \ll \epsilon_{+}(t) - \epsilon_{-}(t) = \sqrt{\Omega(t)^2+\Delta(t)^2}.
\label{eq-adcond}
\ee
When this inequality holds the transitions between the adiabatic states is negligible.
The condition for such adiabatic evolution is expressible as the requirement of a large pulse area, $\calA \gg 1$.

For the special time dependences of a Gaussian pulse shape and a linear chirp, Eqs.~\eqref{Gaussian} and \eqref{chirp}, the adiabatic condition reads \cite{Vasilev2005}
\be\label{adbcond-gs}
\Omega_0\sqrt{2} > \beta \gg \frac2T.
\ee
When the adiabatic condition is fulfilled the AF method delivers efficient population transfer which is robust to variations in the experimental parameters.

\subsection{Shortcuts to adiabaticity (STA)}

Numerous researchers have suggested that  adiabatic passage  might be more effective if the nonadiabatic coupling could be diminished.
The required excitation pulses are designed  by adding to the Hamiltonian a term that cancels the off-diagonal elements of  $\Ha\t$ --- an interaction alteration that has been termed {\em ``shortcut to adiabaticity''} (STA) \cite{STA}. Some studies on the performance of the method in the presence of errors have been also performed \cite{STA-error}.

To achieve the shortcut, consider the Hamiltonian
\be\label{HamiltonianShortcut}
\H'(t) = \frac{\hbar}{2} \bM
	-\Delta(t) & \Omega(t) + \ii \Omega_s(t) \\  \Omega(t)- \ii \Omega_s(t) & \Delta(t)
\eM .
\ee
The additional term $\ii \Omega_s(t) = 2\ii \dot{\theta}(t)$ is chosen exactly equal to the nonadiabatic coupling (but with opposite sign), so that in the adiabatic basis of the \emph{original} Hamiltonian, the transformed shortcut Hamiltonian $\H'(t)$ becomes
\be\label{HamiltonianAdiabatic2}
\Ha'(t) = \hbar  \bM
	\epsilon_{-}(t) & 0 \\  0 & \epsilon_{+}(t)
\eM .
\ee
This scheme, at least in theory, achieves perfect population transfer, because any transition in the adiabatic basis of the \emph{original} Hamiltonian is suppressed.
(Note that the ensuing evolution driven by the new Hamiltonian is nonadiabatic because it proceeds along an eigenstate of the original Hamiltonian which is not an eigenstate of the new Hamiltonian; hence the term ``shortcut to adiabaticity'', albeit established, is inappropriate.)

However, this technique suffers from one essential limitation.
The additional term $\Omega_s(t)$ in the Hamiltonian depends on both $\Omega\t$ and $\Delta\t$.
Therefore, any deviation in these parameters have to be incorporated in the extra term.
However, if such deviations in $\Omega\t$ and $\Delta\t$ are beyond our control, e.g. if they are stochastic, than the shortcut term $\Omega_s(t)$ will deviate from the desired prescription $\Omega_s(t) = 2 \dot{\theta}(t)$.

Moreover, as it turns out, the pulse area of this shortcut term is exactly equal to $\pi$.
Indeed, the nonadiabatic coupling for the Gaussian pulse shape \eqref{Gaussian} and the linear chirp \eqref{chirp} reads
\be
\dot{\theta} = -\frac{\Omega_0 (\beta/T)\e^{(t/T)^2}(2t^2+T^2)}{2(\beta^2 t^2\e^{2(t/T)^2}+\Omega_0^2 T^2)} ,
\ee
see Fig.~\ref{fig:shapes}.
One can easily show that $\int_{-\infty}^{\infty} dt \, \dot{\theta} = -\pi/2$.
Hence the extra term $\Omega_s(t) = 2\dot{\theta}(t)$ in the Hamiltonian \eqref{HamiltonianShortcut} has an area of exactly $\pi$.
Obviously, if one is able to produce a pulse with a pulse area of $\pi$, it could be used directly for resonant excitation.
Nevertheless, here we examine the performance of the STA method to find out if it is superior to resonant and adiabatic excitation and worth the effort to produce an additional shaped field.

When we examine the performance of the STA technique, in order to be fair we need to fix the $\Omega_0$ parameter in the expression for $\dot{\theta}$.
Leaving it free would lead to a spurious perfect robustness of the technique.
We will denote the fixed value of $\Omega_0$ in $\dot{\theta}$ as $\Omega_0^a$. The same is valid for $\beta$ and $T$, whose values in the shortcut term we denote with $\beta^a$ and $T^a$.

We also note that, the shortcut to adiabaticity, as implemented in Eq.~\eqref{HamiltonianShortcut}, is not the only possible approach. It is well known that a complex Rabi frequency is equivalent to a real Rabi frequency and additional detuning. Such physical realization of the shortcut may lead to different performance of the method. However, our simulations have shown that both approaches produce very similar results and therefore we only present the one with the additional field in the Rabi frequency.

%

\subsection{Shaped pulses (SP)}

The ``single-shot shaped pulse'' technique was recently introduced by Gu\'erin and co-authors \cite{SingleShot}.
We shall refer to it as ``shaped pulse'' (SP) in order to avoid confusion with RE and AF (and even STA in some sense), which also use single pulses, and also to emphasize the fact that the shapes of the Rabi frequency and the detuning are used as controls.
By suitably choosing the temporal shape of the Rabi frequency and the detuning one can obtain robust population transfer with a moderate pulse area.
This is achieved by a procedure of nullification of integrals, obtained after a perturbative expansion.
The theoretical details of the method are thoroughly explained in Refs.~\cite{SingleShot}.
The explicit analytic time dependence of the Rabi frequency and the detuning for SP can be derived from the following equations,
\bse
\begin{align}
&\Omega(t) = \dot{\gamma}(t)\frac{\sin\theta(t)}{\cos\phi(t)},\\
&\Delta(t) = \dot{\phi}(t)-\Omega(t)\cos\phi(t)\cot\theta(t),
\end{align}
\ese
where
\bse
\begin{align}
\theta(t) &= \frac{\pi}{2}[\text{erf}(t/T)+1],\\
\gamma(t) &= 2\theta(t) + C_1 \sin 2\theta(t) + \cdots + C_n \sin 2n\theta(t),\\
\phi(t) &= \arctan\left[\frac{\dot{\theta}(t)}{\sin\theta(t)\dot{\gamma}(t)}\right].
\end{align}
\ese
The first few coefficients $C_n$ need to be chosen such that to nullify the required integrals in the perturbative expansion.
In the simulations below, we use the type-$A$ SP of order 7 \cite{SingleShot}, which we denote as $A 7$ and for which we have $C_1=-3.46$, $C_2=-1.365$, $C3 = -0.5$.
All higher-order coefficients $C_n$ are equal to zero.
For visualization purpose, we  plot the Rabi frequency and the detuning for this SP in Fig.~\ref{fig:shapes}.
As seen from Fig.~\ref{fig:shapes}, careful shaping of both the Rabi frequency and the detuning is needed for this method.

\subsection{Composite pulses}

A well-studied method for creating quantum changes that combines a robustness akin to adiabatic techniques with a high fidelity akin to resonant excitation uses a sequence of pulses, cf. Fig.~\ref{fig:shapes},
\be\label{CP}
A_{\phi_1} A_{\phi_2} \cdots A_{\phi_N},
\ee
whose phases $\phi_k$ are carefully crafted to produce a desired result, a structure known as {\em composite pulses} (CP).
The relative phases serve as the control parameters, to be adjusted to produce some specific objective: increased robustness to parameter errors is obtained for one set of composite phases while increased sensitivity and selectivity is obtained for another set of phases.
One can construct composite pulses with a variety of excitation profiles: broadband, narrowband and passband profiles, or even robust coherent superpositions and quantum gates.
The pulse areas in the sequence \eqref{CP} are assumed equal, as it is the case for many CPs, but they may differ from pulse to pulse and be used as control parameters too.
For each of these pulses a phase shift in the Rabi frequency  $\Omega(t) \to \Omega(t)\e^{\ii \phi}$ is imprinted onto the off-diagonal elements of the propagator \eqref{Propagator} as
\be
\U(\phi) = \bM a & b \e^{\ii \phi} \\ - b^\ast \e^{-\ii \phi} & a^\ast \eM.
\ee
The propagator of a composite sequence of $N$ such identical pulses with phases $\phi_k$ $(k=1,2,\ldots,N)$ has the form of a product,
\be
\U^{(N)} = \U(\phi_N)\U(\phi_{N-1}) \ldots \U(\phi_2)\U(\phi_{1}).
\ee
The latter expression for the total propagator is used to generate a set of equations for the composite phases, depending on the desired excitation profile.

We have used two types of CPs for our comparison.
The first is the composite adiabatic passage (denoted by CAP) technique, which uses a sequence of chirped pulses, and in such a way is benefiting from the symmetry properties of the Hamiltonian to further improve adiabaticity \cite{Torosov2011PRL,Schraft2013}.
The second is the universal CPs (denoted by UCP) which cancel errors versus any experimental parameter \cite{GenovUniversal}.

As a representative of the CAP technique \cite{Torosov2011PRL} we use the shortest three-pulse sequence of the family,
\be\label{CAP3}
C3 = A_0 A_{\frac23 \pi} A_0.
\ee
As a representative of the universal CP family, we use the five-pulse sequence \cite{GenovUniversal}
\be\label{U5}
U5 = A_0 A_{\frac56 \pi} A_{\frac13 \pi} A_{\frac56\pi} A_0.
\ee
For either CPs we assume the Gaussian pulse shapes of Eq.~\eqref{Gaussian}.
For the universal CPs we assume exact resonance in the errorless case, which for CAP we assume the linear chirp of Eq.~\eqref{chirp}.


\section{Experimental errors}

We next examine how, for the quantum control protocols discussed above, the effectiveness of population swapping is affected by   experimental errors, originating from specific experimental implementations and conditions.
We consider several types of experimental errors, each of which occurs in particular types of control.

\subsection{Rabi frequency error}

\begin{figure}[tb]
	\includegraphics[width=8.cm]{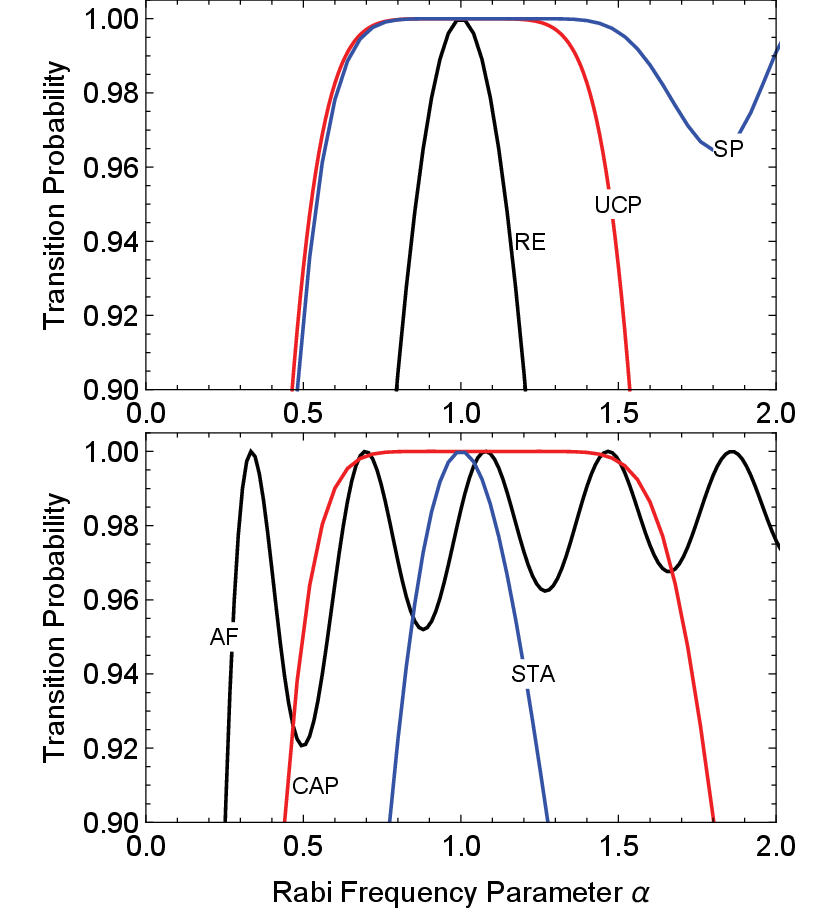}
	\caption{Transition probability vs the Rabi frequency parameter $\alpha$, see Eq.~\eqref{alpha}, for resonant excitation (RE), universal composite pulses (UCP), adiabatic following (AF), shortcuts (STA), composite adiabatic passage (CAP), and shaped pulse (SP).
The values of the parameters are $\Omega_0=\sqrt{\pi}/T$ for RE, UCP, CAP and STA, and $\Omega_0=5\sqrt{\pi}/T$ for AF, with $\beta=4/T$ and $\Omega_0^a = \Omega_0$ for STA. For CAP, $\beta=1/T$.
}
	\label{vsOmega1}
\end{figure}

We first consider errors that affect the Rabi frequency.
Such errors may occur in physical situations of atoms or ions in a non-uniform spatial distribution of an external laser, microwave or radiofrequency (rf) field.
Then the atoms in the edge are subject to a field which is only a fraction of the field which affects the atoms in the centre.
Such a variation can occur in a dense cloud of ultracold atoms in a dipole trap or atoms in a cell; then a tightly focused laser beam has some variation of its spatial intensity distribution over the atomic ensemble.
Another example is coherent manipulation of doped solids with a typical crystal size of a few mm and a typical variation of the Rabi frequency of some 20-30\% over the crystal.
In yet another example, when an atomic beam crosses a focused laser beam, in a crossed-beam scenario, only atoms passing through the center of the laser beam will experience the maximum laser intensity, whereas off-center atoms will ``see'' reduced intensity.
We note that the intensity of a laser source may vary in time, as it typically decreases over the course of the experiment.

\begin{figure}[tb]
	\includegraphics[width=8.cm]{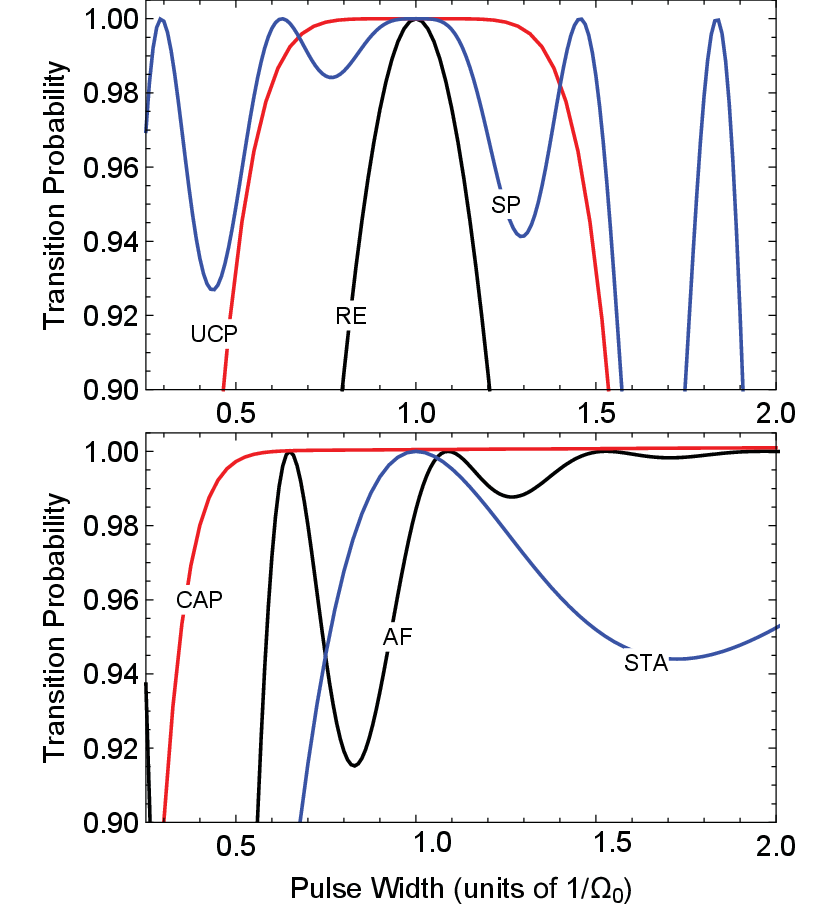}
	\caption{Transition probability vs pulse width for resonant excitation (RE), universal composite pulses (UCP), adiabatic evolution (AF), shortcuts (STA), composite adiabatic passage (CAP), and shaped pulse (SP).
The values of the parameters are $\Omega_0=\sqrt{\pi}/T$ for RE, UCP, CAP and STA, and $\Omega_0=5\sqrt{\pi}/T$ for AF, with $\beta=4/T$ and $T^a = \sqrt{\pi}/\Omega_0$. For CAP, $\beta=1/T$.
}
	\label{vsTime}
\end{figure}

To quantify uncontrolled and hence error-inducing intensity variation we introduce a dimensionless parameter $\alpha$, writing  the Rabi frequency as
\be\label{alpha}
\Omega(t) \to \alpha \Omega(t).
\ee
For $\alpha=1$, we have the default case for each technique, while deviations from this value account for Rabi frequency errors.
The value of unity for  $\alpha$ corresponds to a different pulse area $\calA$ for the six techniques: $\calA=\pi$ for RE; $\calA=3\pi$ for AF and STA, $\calA=3.86\pi$ for SP, $\calA=5\pi$ for UCP and CAP.
We will introduce such dimensionless factors to measure variations in other experimental parameters as well.

In Fig.~\ref{vsOmega1}, we compare the accuracy of each of the six methods by plotting the transition probability as a function of the Rabi frequency parameter $\alpha$.
We conclude that in the vicinity of $\alpha=1$, the UCP, CAP, and SP techniques behave quite well,
being robust on either sides of $\alpha=1$.
As $\alpha$ increases, the AF technique improves, as it is well known, but does not reach unity.
Again, as it is in the textbooks, RE method is sensitive to variations in the pulse area.
The STA method achieves an efficiency close to the RE technique.

\label{transit-time-sec}
\subsection{Pulse duration error}


Pulse duration errors can in principle occur due to the pulsed field source, but they are usually well controlled.
However, there are experimental situations when they are unavoidable.
Consider a situation where an atomic beam passes across a laser beam.
The atoms have some distribution over their longitudinal velocities, which leads to a distribution of transit times.
Therefore, each atom may experience a different interaction duration, i.e. a different pulse width.

We plot in Fig.~\ref{vsTime} the transition probability versus $T$ and find that the effect of the variation in the pulse width is similar to that of the Rabi frequency variation, with some twists.
The CAP technique outperforms its competitors, followed by UCP.
Again, the RE method is sensitive to variations of $T$, while the AF improves as $T$ increases, illustrating well known features of the two methods.
The SP and STA  methods maintain values above 90\% over broad ranges of $T$ but fail to keep the probability at ultrahigh efficiencies above 99.9\%, thereby underperforming CAP and UCP.

\begin{figure}[tb]
	\includegraphics[width=0.5\textwidth]{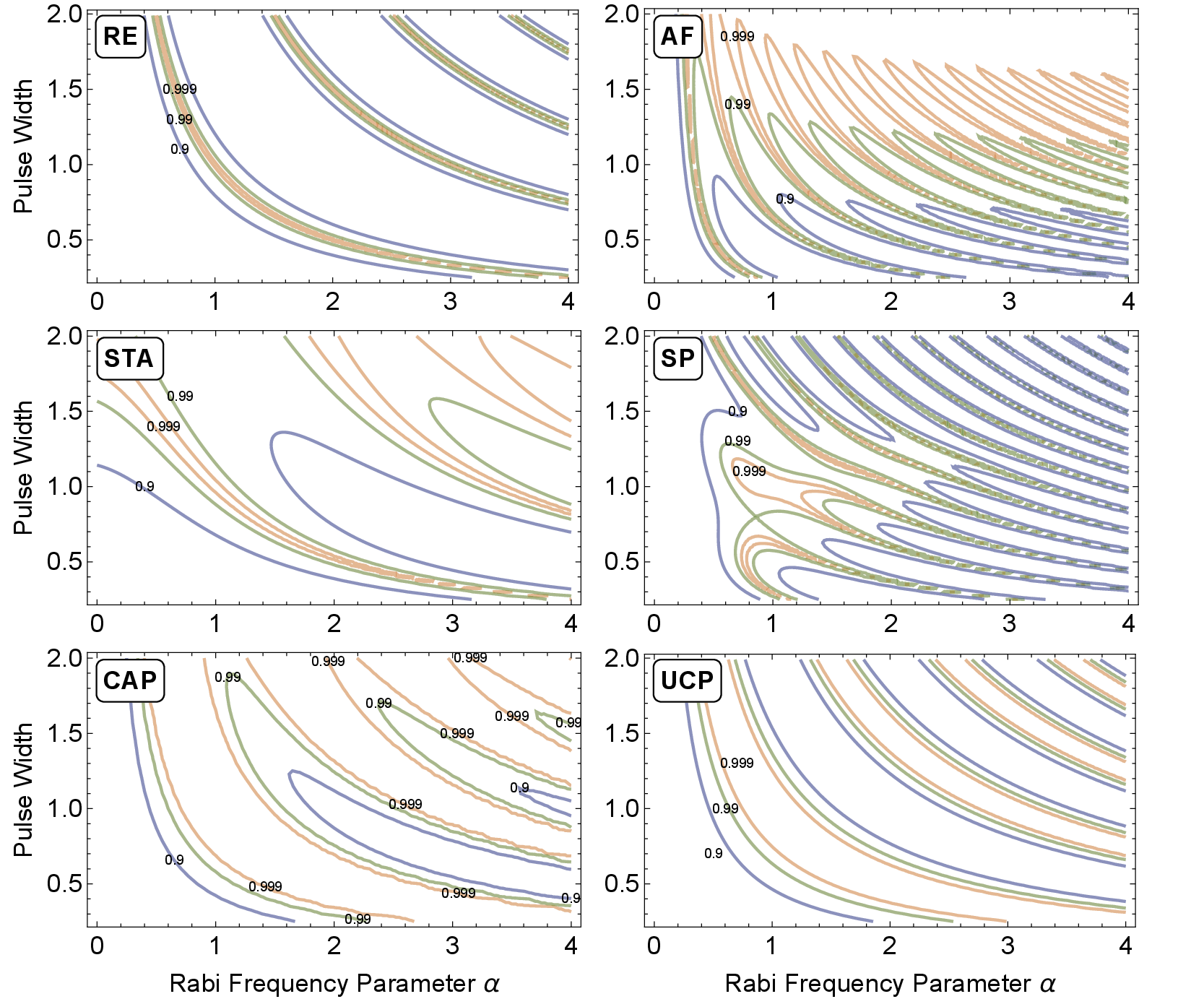}
	\caption{Contours of transition probability vs the Rabi frequency parameter $\alpha$, see Eq.~\eqref{alpha}, and the pulse width (in units $1/\Omega_0$ for resonant excitation (RE), universal composite pulses (UCP), adiabatic following (AF), shortcuts (STA), composite adiabatic passage (CAP), and shaped pulse (SP).
The values of the parameters are $\Omega_0=\sqrt{\pi}/T$ for RE, UCP, STA and CAP, and $\Omega_0=5\sqrt{\pi}/T$ for AF, with $\beta=4/T$ and $\Omega_0^a = \Omega_0$. For CAP, $\beta=1/T$.\\
}
	\label{vsOmega2D}
\end{figure}

In order to gain a broader view and further insight, in Fig.~\ref{vsOmega2D}, we compare the accuracy of each of the six methods by plotting the transition probability as a function of both the Rabi frequency parameter $\alpha$ and the pulse duration $T$.
Because the pulse area is proportional to the product of these parameters, the RE method shows oscillatory behavior with tiny high-efficiency regions, while the AF method steadily improves as $\alpha$ and especially $T$ increase.
The SP method has moderately broad ranges around the point $\{ \alpha=1,\ T=1/\Omega_0 \}$ but reduced performance elsewhere.
The UCP and CAP methods feature broad ranges of high efficiency not only around the point $\{ \alpha=1,\ T=1/\Omega_0 \}$ but over broad strips; they are the clear winners here.

\subsection{Detuning errors}

\begin{figure}[tb]
	\includegraphics[width=8.cm]{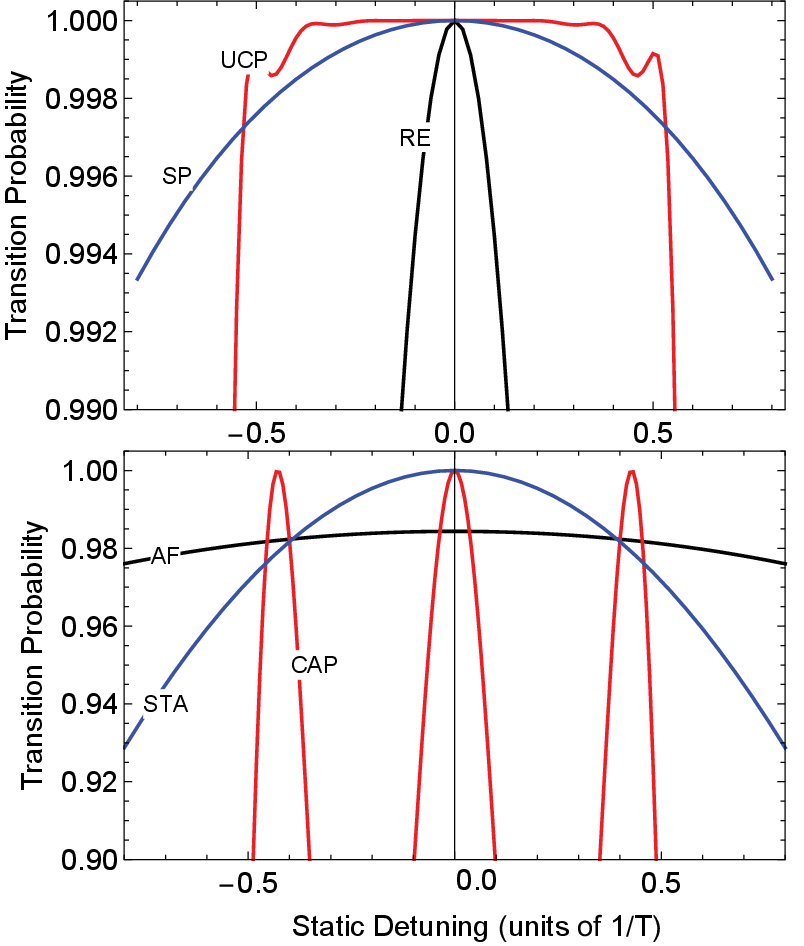}
	\caption{Transition probability vs static detuning $\delta$ for resonant excitation (RE), universal composite pulses (UCP), adiabatic evolution (AF), shortcuts (STA), composite adiabatic passage (CAP), and shaped pulse (SP).
The values of the parameters are $\Omega_0=\sqrt{\pi}/T$ for RE, UCP, CAP and STA, and $\Omega_0=5\sqrt{\pi}/T$ for AF, with $\beta=4/T$. For CAP, $\beta=1/T$.
Note the different vertical scales in the two frames.
}
	\label{vsDelta}
\end{figure}

An important experimental error is the detuning error.
Because the detuning is the difference between the Bohr transition frequency $\omega_0$ and the driving field frequency $\omega$, errors in the detuning can be introduced by either of these frequencies.

The Bohr transition frequency can be altered by uncompensated stray electric or magnetic fields.
It can also suffer from inhomogeneous broadening.
For example, in experiments with optical memories in doped crystals \cite{Schraft2013}, the hyperfine transition of the ions in the crystal is broadened due to the inhomogeneity in the medium, i.e. each ion ``sees'' a different environment and therefore, experiences a different energy level shift.
This leads to a different detuning in each ion, and the overall signal, as an average of the signals from all ions, will suffer from this inhomogeneous broadening.


\begin{figure}[tb]
	\includegraphics[width=0.5\textwidth]{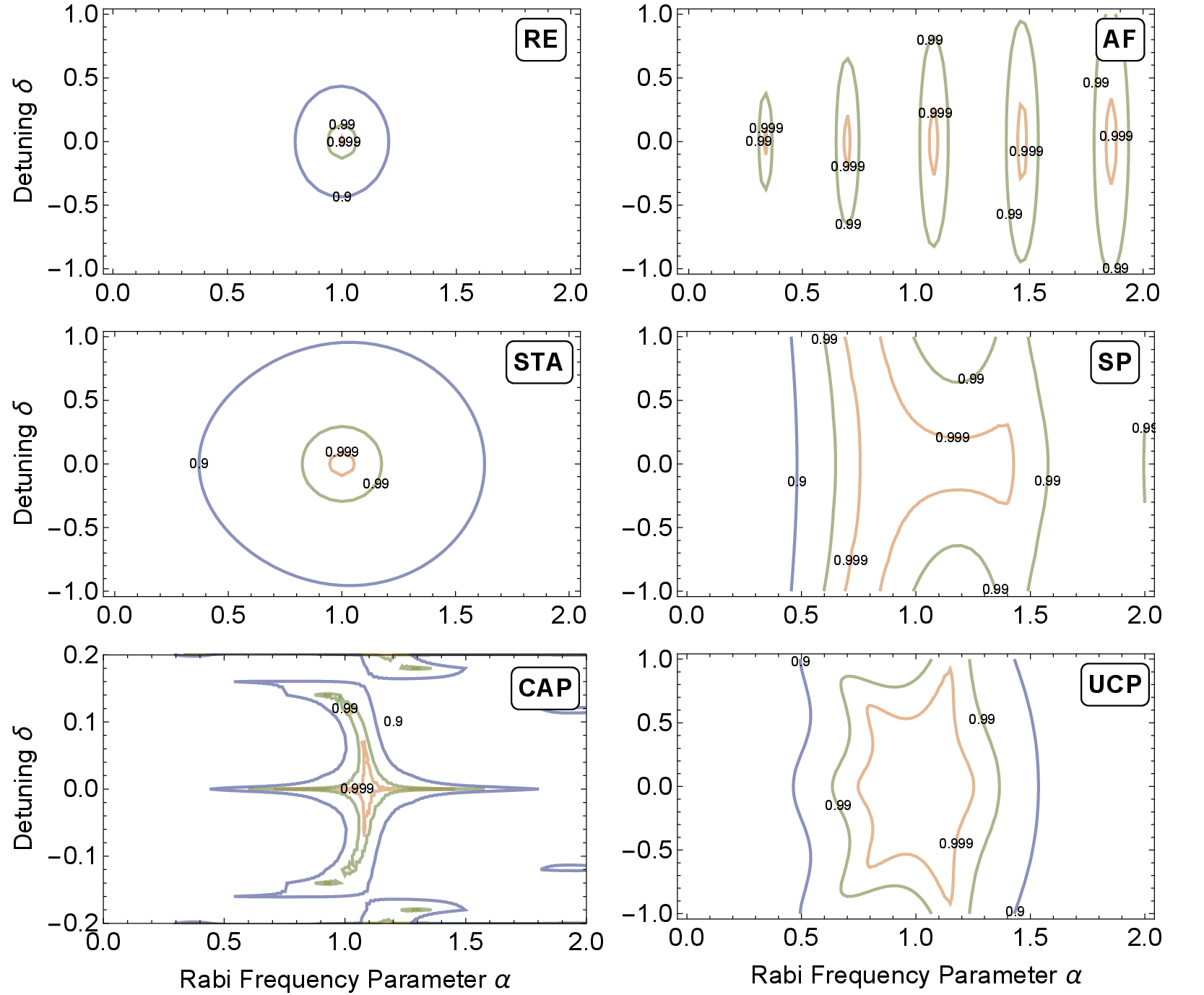}
	\caption{Contours of transition probability vs control parameters $\alpha$ and $\delta$ (in units $1/T$) for resonant excitation (RE), universal composite pulses (UCP), adiabatic following (AF), shortcuts (STA), composite adiabatic passage (CAP), and shaped pulse (SP). The values of the parameters are $\Omega_0=\sqrt{\pi}/T$ for RE, UCP, CAP and STA, and $\Omega_0=5\sqrt{\pi}/T$ for AF, with $\beta=4/T$. For CAP, $\beta=1/T$.
		Note the different vertical scale in the CAP frame.
}
	\label{vsDA}
\end{figure}

The driving frequency of the external field may suffer from unwanted uncertainties too.
Obviously, the frequency of the radiation source itself, whether laser, microwave or rf, may fluctuate during the course of measurement, thereby generating a random detuning.
Furthermore, as discussed in Sec.~\ref{transit-time-sec}, an atomic beam, interacting with a laser field,  may suffer from the distribution of the velocities of the atoms.
While the longitudinal velocity spread would result in transit-time broadening, the transverse velocity distribution would lead to Doppler broadening, and therefore to a detuning error.


In order to study the stability of the transition probability versus the detuning offset we add a parameter $\delta$ to the detuning in the Hamiltonian,
\be
\Delta(t) \to \Delta(t) + \delta,
\ee
which we call \emph{static detuning}.

The stability of the transition probability versus the static detuning is illustrated in Fig.~\ref{vsDelta}.
In this case, the UCP method outperforms the other techniques, followed by SP and then STA.
The CAP technique is sensitive to static detuning because it is based on certain symmetries of the control functions which are broken by the static detuning.
As usual, the RE method is sensitive too, while the AF method is stable versus variations of $\delta$ but at lower efficiencies than the other techniques.

These conclusions are reaffirmed in the 2D plot in Fig.~\ref{vsDA}, where both errors in the detuning and in the peak Rabi frequency are present.
The UCP and SP techniques are the clear winners, followed by the STA and CAP methods which feature some robustness but at much lower magnitude compared to UCP and SP.

\subsection{Chirp error}

Next, we consider the presence of an undesirable chirp in the field.
Such a chirp may occur, e.g. in ultrafast excitation by picosecond and femtosecond pulses because the method of generation of such pulses (stretching, amplification and compression) introduces a chirp and special care is needed to remove it \cite{Wollenhaupt}.
An unwanted chirp can be present in microwave and rf fields too \cite{Zarantonello-thesis}.

\begin{figure}[tb]
	\includegraphics[width=8.cm]{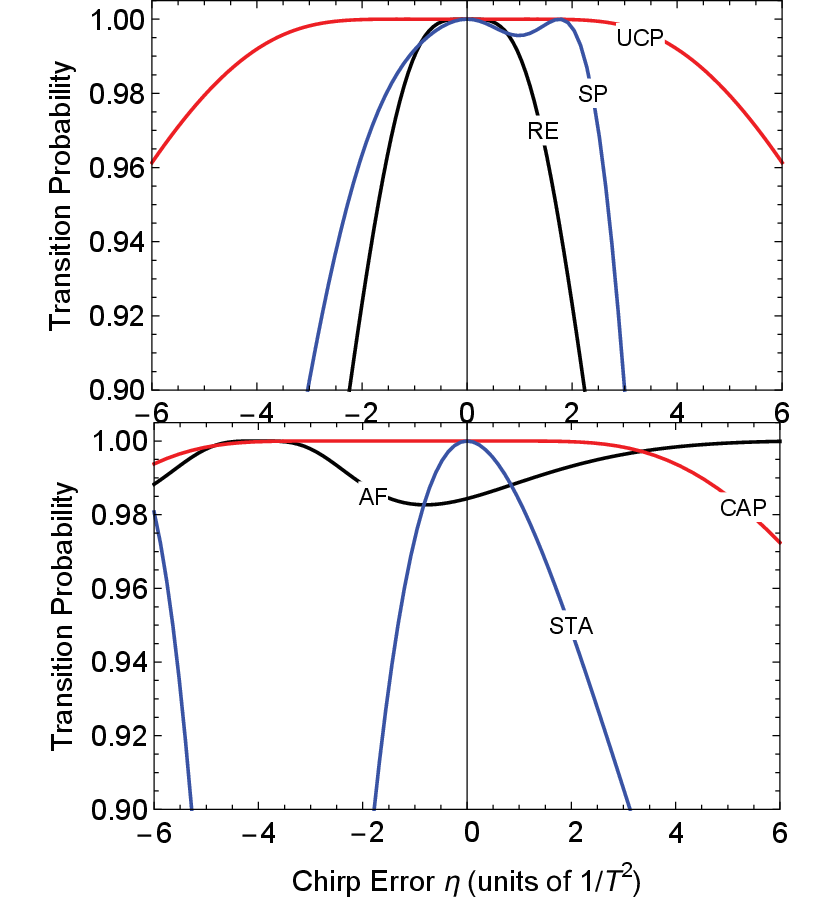}
	\caption{Transition probability vs chirp error $\eta$ for resonant excitation (RE), universal composite pulses (CP), adiabatic evolution (AF), shortcuts (STA), composite adiabatic passage (CAP), and shaped pulse (SP).
The values of the parameters are $\Omega_0=\sqrt{\pi}/T$ for RE, UCP, CAP and STA, and $\Omega_0=5\sqrt{\pi}/T$ for AF, with $\beta=4/T$ and $\beta^a=4/T$. For CAP, $\beta=1/T$. \\
	}
	\label{vsChirp}
\end{figure}

In order to study the stability of the transition probability versus unwanted chirp we add a parameter $\eta t$ to the detuning in the Hamiltonian,
\be
\Delta(t) \to \Delta(t) + \eta t.
\ee
The effect of this unwanted additional chirp on the transition probability is illustrated in Fig.~\ref{vsChirp}.
We find that UCP and CAP outperform the other methods.
This is expected because the UCP method is robust to any parameter error, while the CAP method uses chirp (to which it is robust) and adding more chirp does not make much difference as long as this additional chirp does not exceed the original one.
A positive additional chirp in the AF makes the method more adiabatic, which increases the fidelity. A small negative chirp error makes the AF method equivalent to the RE and again increases the fidelity, as long as the rror is small, which is expected.
Larger chirp deteriorates the adiabatic condition \eqref{adbcond-gs} and reduces the efficiency, which is not in the range shown on Fig.~\ref{vsChirp}.
Notably, even resonant pulses behave well for some range, because the unwanted chirp turns the RE method into the AF method and brings some robustness to the RE method.
The worst performer is STA, followed by SP, because the fine balance between the main and shortcut fields (STA), or between the shapes of the control functions (SP) is destroyed by the chirp.

\subsection{Shape error}

\begin{figure}[tb]
	\includegraphics[width=8cm]{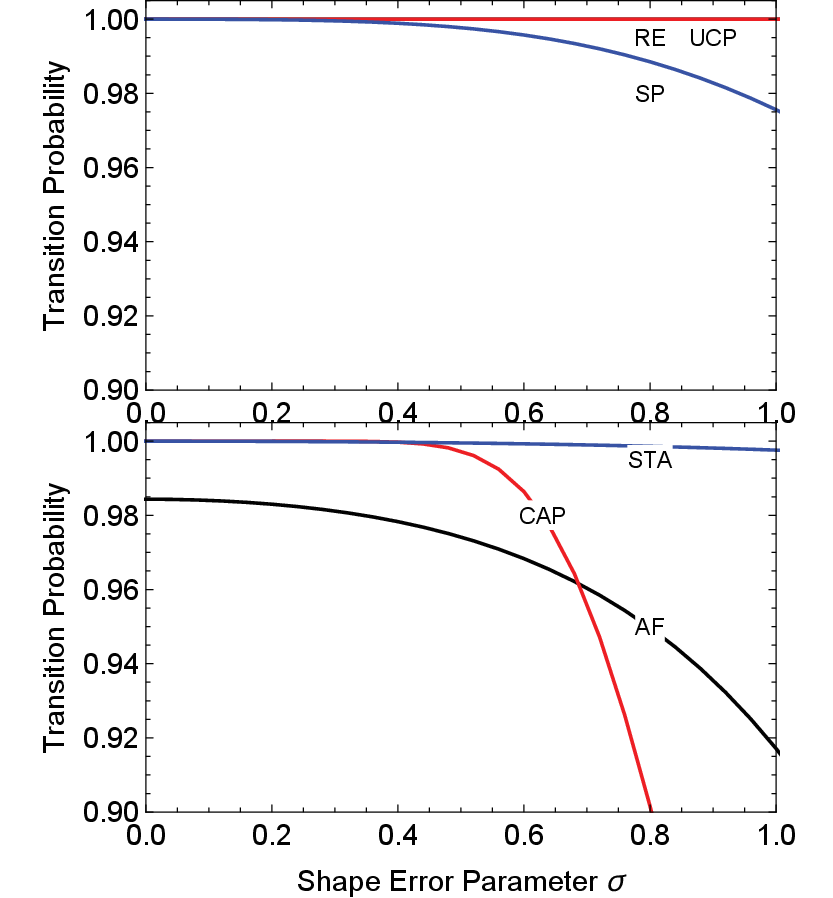}
	\caption{Transition probability vs shape-error parameter for resonant excitation (RE), universal composite pulses (UCP), adiabatic evolution (AF), shortcuts (STA), composite adiabatic passage (CAP), and shaped pulse (SP).
The parameter values are $\Omega_0=\sqrt{\pi}/T$ for RE, UCP, STA and CAP, and $\Omega_0=5\sqrt{\pi}/T$ for AF, with $\beta=4/T$. For CAP, $\beta=1/T$.\\
}
	\label{vsShape}
\end{figure}

Finally, we consider the presence of an error in the shape of the pulse.
We model this type of error by adding an additional term to the original Gaussian shape,
\be
\Omega(t) \to \Omega_0 [1+ \sigma \tanh(t/T)],
\ee
where $\sigma$ is a parameter, which measures the strength of the error in our pulse shape.
We have chosen the antisymmetric tanh function in order not to change the pulse area but only the shape of the driving pulse; otherwise there would be an ambiguity if the change is caused by the different pulse shape or the different pulse area.

We can see from Fig.~\ref{vsShape} that the RE and UCP methods are not affected at all by the pulse shape, as it should be the case, because the resonant probability depends on the pulse area only.
The STA method is virtually unaffected, with very little dependence of the shape, which is hard to be seen on the figure.
The efficiency of the SP method, which relies upon pulse shaping, naturally deteriorates as the pulse shape error grows. Finally the CAP method, which relies on certain symmetries in the field parameters, is only robust for small to moderate deviations in the pulse shape.

\subsection{RWA errors}

So far we have assumed that the RWA applies.
This may not be the case, e.g. in the so-called strong-coupling regime when the Rabi frequency is comparable to the transition frequency.
Then the counter-rotating term, discarded in the RWA, may have a significant influence on the dynamics.
We point out that in laser-atom interaction, the counter-rotating term emerges when the qubit transition is driven by linearly polarized light \cite{Shore1990}.
If it is driven by circularly polarized light then no such term is present \cite{Shore1990}.

\begin{figure}[tb]
	\includegraphics[width=8.cm]{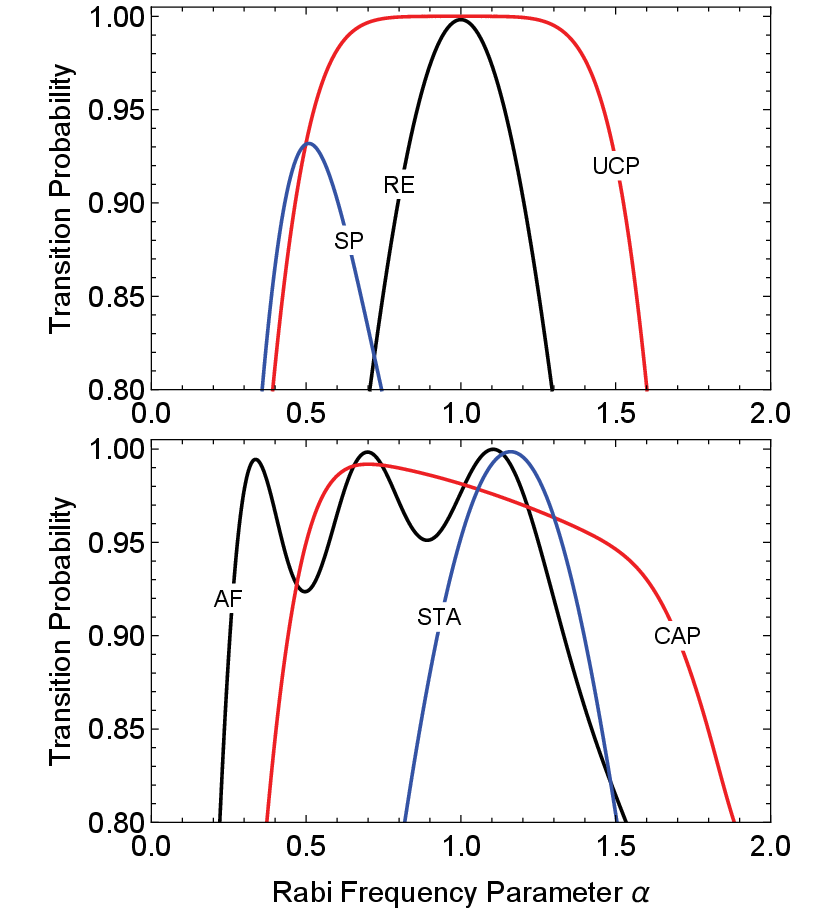}
	\caption{Same as Fig.~\ref{vsOmega1}, but without RWA.
The simulations are conducted by using the Hamiltonian of Eq.~\eqref{Hamiltonian-noRWA}.
The parameters in the counter-rotating term are chosen as $\omega_{0}=5\sqrt{\pi}/T$ and $\omega=\omega_0+\Delta(t)$, where $\omega_{0}$ is the Bohr transition frequency, $\omega$ is the frequency of the external field, and $\Delta(t)$ is the same detuning as the one used in Fig.~\ref{vsOmega1}.}
	\label{vsOmegaNoRWA}
\end{figure}

To this end, we have conducted simulations without the RWA.
In the interaction representation, we have the Hamiltonian
\begin{align}
\H(t) &= \hbar \Omega(t) \cos \eta(t)\,e^{-i \omega_0 t} \ket{1} \bra{2} + h.c., \notag \\
 &= \half \hbar \Omega(t)\, [e^{-i [\omega_0 t - \eta(t) ] } + e^{ -i [\omega_0 t + \eta(t) ] } ] \ket{1} \bra{2} + h.c.,
\label{Hamiltonian-noRWA}
\end{align}
where $\eta(t) = \int^t \omega(t') dt'$ and hence $\partial_t \eta(t) = \omega(t)$
\footnote{ Here we have assumed that the temporal variation of the detuning $\Delta(t) = \omega_0 - \omega(t)$, needed as a control parameter in several of the control techniques we consider, derives from the temporal variation of the frequency $\omega(t)$ of the driving field.
Similar results are obtained when the temporal variation is incorporated in the Bohr transition frequency $\omega_0$. }.
The first term in Eq.~\eqref{Hamiltonian-noRWA} generates the time-dependent detuning and the second term is the counter-rotating one, which is neglected in the RWA.
However, the RWA is justified only when the sum $\omega_0 + \omega(t)$ is much greater than the Rabi frequency $\Omega(t)$.
Here we have retained the counter-rotating term and the results of the numerical simulations are shown in Fig.~\ref{vsOmegaNoRWA}.
The UCP technique outperforms all other techniques by a wide margin.
The CAP technique retains some robustness to errors but at reduced efficiency.
The other techniques can provide high efficiency for specific Rabi frequency values only.
The SP method is the worst performer here as it is neither accurate nor robust.

\section{Discussion and Conclusions}

\begin{table*}[tb]
\begin{tabular}{|c|c|cccccccc|}
\hline
 Technique         & Rabi & Duration & Detuning & Chirp & Shape & Phase & no-RWA & Speed & Accuracy \\
 & frequency &  &  &  &  &  &  & &  \\
 \hline
RE & sensitive & sensitive & sensitive & moderate & robust & robust & sensitive & fast & high \\
AF & moderate & robust & robust & robust & moderate & robust & sensitive & slow & moderate \\
STA  & sensitive & moderate & moderate & sensitive & robust & robust & sensitive & moderate & high \\
SP & robust & moderate & robust & moderate & moderate & robust & sensitive & moderate & high \\
CAP & robust & robust & moderate & robust & moderate & sensitive & moderate & moderate & high \\
UCP & robust  & robust & robust & robust & robust & sensitive & robust & moderate & high \\
\hline
\end{tabular}
\caption{Summary of features of the six coherent control techniques studied here.
The resilience to experimental errors in various control parameters is characterized by the terms ``sensitive'', ``moderate'' and ``robust'', and the speed and the accuracy are given in the last two columns.}
\label{table}
\end{table*}

In the present work we compared six popular techniques for coherent control of two-state quantum systems, by simulating their behavior in real experimental situations.
We tested the accuracy and robustness of resonant excitation, adiabatic evolution, shortcuts to adiabaticity, universal composite pulses, composite adiabatic passage, and shaped pulses in situations where errors in the driving field intensity and duration, the detuning, the chirp, the pulse shape, and the pulse phase are present.
Our studies have focused in particular on the ability of the six control techniques to achieve the ultrahigh efficiency suitable for quantum computation.

Table \ref{table} summarizes the observations of the performance of these six control techniques.

The properties of resonant excitation and adiabatic following are well known and are used here for benchmarking purposes \cite{Shore1990}.
\emph{Resonant excitation} is fast and accurate but it is sensitive to errors in the pulse amplitude, duration and detuning; however, it is insensitive to variations in the pulse shape and phase, and only moderately sensitive to a possible chirp as it turns resonant excitation into adiabatic following.
\emph{Adiabatic following} is robust to errors in all experimental parameters but it is slow and does not deliver ultrahigh accuracy.

The shaped-pulse approach is robust to amplitude, detuning and phase errors but not so much to duration, chirp, shape and RWA errors.
This is not surprising because the precise shapes of the Rabi frequency and the detuning are essential for the method.
The shortcut to adiabaticity approach features similarly to shaped pulses but it performs worse to amplitude, detuning and chirp errors.

The two composite-pulse techniques are very robust to parameter errors except the phases in the pulse sequence, which is expected because they are the control parameters.
While the universal composite pulses are very robust to errors in all experimental parameters, the composite adiabatic passage approach is sensitive to static detuning and RWA errors because the latter violate the symmetry assumption used to derive the sequence.

The universal composite pulse emerges as the best overall performer as it holds up very well against almost all considered errors, even beyond the RWA.
However, one should bear in mind that it requires a relatively large total pulse area of $5\pi$ and accurate control of the phases of the constituent pulses.
The CAP technique requires pulse area of $5\pi$ and behaves similarly to the universal composite pulses except for detuning errors.
The shaped pulse requires a pulse area of nearly $4\pi$ and performs well, but not extremely well, against most sorts of errors:
 it features significant robustness to errors at the level of 99\% efficiency but less robustness at the 99.9\% and higher, toward the quantum computation benchmark values, e.g. against the pulse duration and the pulse shape errors.
A similar conclusion as for the shaped pulse can be drawn for the shortcut approach.
We should also mention that the shortcut technique is sometimes promoted as being faster than alternative methods except for resonant excitation.
This is due to the eliminated need of (slow) adiabatic evolution.
However, one could argue that the pulse area of the resonant shortcut field is equal to $\pi$ and it alone can produce complete population transfer.
Therefore, the original field for which the shortcut is introduced becomes redundant and without it, the shortcut is reduced to a resonant $\pi$ pulse.
Single-shot pulses and shortcut to adiabaticity are worth considering, especially if pulse shaping can be produced with high accuracy.
Yet, achieving very high population transfer efficiency in broad parameter ranges may be difficult to achieve with these last two methods.



Finally, we note that there are other pulse choices available for some of the methods (e.g. SP, CAP, UCP)  discussed here, which may perform better against some types of the errors (but worse against other types).
We have restricted our present study to single representatives of each technique which, in our view, deliver the best cost-to-feature balance in the ultrahigh efficiency regime.

\acknowledgments
This work is supported by the European Commission's Horizon-2020 Flagship on Quantum Technologies project 820314 (MicroQC).


\end{document}